# Laboratory Measurement of Volatile Ice Vapor Pressures with a Quartz Crystal Microbalance

**W.M. Grundy[1,2], S.C. Tegler[2], J.K. Steckloff[3,4], S.P. Tan[3], M.J. Loeffler[2,5], A.V. Jasko[6,2], K.J. Koga[1,2], B.P. Blakley[2,7], S.M. Raposa[2], A.E. Engle[2], C.L. Thieberger[2], J. Hanley[1,2], G.E. Lindberg[8,5], M.D. Gomez[2], and A.O. Madden-Watson[2]**

1. Lowell Observatory, Flagstaff Arizona.
2. Department of Astronomy and Planetary Science, Northern Arizona University, Flagstaff Arizona.
3. Planetary Science Institute, Tucson Arizona.
4. Department of Aerospace Engineering and Engineering Mechanics, University of Texas at Austin, Austin Texas.
5. Center for Material Interfaces in Research and Applications, Northern Arizona University, Flagstaff Arizona.
6. Case Western Reserve University, Cleveland Ohio.
7. Pasadena City College, Pasadena California.
8. Department of Chemistry and Biochemistry, Northern Arizona University, Flagstaff Arizona.

## Abstract

Nitrogen, carbon monoxide, and methane are key materials in the far outer Solar System where their high volatility enables them to sublimate, potentially driving activity at very low temperatures. Knowledge of their vapor pressures and latent heats of sublimation at relevant temperatures is needed to model the processes involved. We describe a method for using a quartz crystal microbalance to measure the sublimation flux of these volatile ices in the free molecular flow regime, accounting for the simultaneous sublimation from and condensation onto the quartz crystal to derive vapor pressures and latent heats of sublimation. We find vapor pressures to be somewhat lower than previous estimates in literature, with carbon monoxide being the most discrepant of the three species, almost an order of magnitude lower than had been thought. These results have important implications across a variety of astrophysical and planetary environments.

## Introduction

Nitrogen, carbon monoxide, and methane occur as solids in diverse low temperature environments at the outer edge of the Solar System. They are thought to be legacies of the chemical inventory from the Solar System's formation via the gravitational collapse of a giant molecular cloud of gas that led to the creation of the proto-Sun and its surrounding, disk-shaped protoplanetary nebula (e.g., Öberg et al. 2011). Frozen $N_2$, CO, and $CH_4$ are constituents of the icy mantles around molecular cloud dust particles. On arrival at the protoplanetary nebula, dust particles may warm sufficiently for these "hypervolatiles" to sublimate and became part of the local nebular gas inventory. In the coldest, outer parts of the nebula, they may remain on or condense back onto dust particles, enabling them to be incorporated into planetesimals, the building blocks of planets (e.g., Ciesla 2009; Krijt et al. 2018). After the Sun's T Tauri winds cleared the residual dust and gas and the Sun fully illuminated its planetesimals for the first time, it drove sublimation of hypervolatiles causing an early sublimative era of widespread cometary activity (e.g., Steckloff et al. 2021) and partially or completely devolatilizing objects as a function of heliocentric distance and other thermal parameters. Larger, planetary-scale bodies with stronger gravity and/or cooler surfaces could better retain their volatiles (e.g., Schaller and Brown 2007; Johnson et al. 2015). In our Solar System, Pluto and Triton host abundant $N_2$, CO, and $CH_4$ on their surfaces, as revealed by the vibrational absorption features of these molecules in near-infrared reflectance spectra (e.g., Owen et al. 1993; Cruikshank et al. 1993). Sublimation of $N_2$, CO, and $CH_4$ powered by sunlight leads to their seasonal migration (e.g., Spencer et al. 1997; Trafton et al. 1998). Regionally distinct combinations of volatile materials and seasonal climate factors create strikingly diverse geological provinces on these bodies (e.g., Stern et al. 2015; Moore et al. 2016). Methane has numerous strong and distinctive near-infrared vibrational absorption bands, making it easiest to detect of the three hypervolatiles. It has also been detected on the surfaces of other bodies, including Eris, Makemake, Quaoar, and Sedna (e.g., Barucci et al. 2005; Licandro et al. 2006; Dumas et al. 2007; Dalle Ore et al. 2009), as well as escaping from the Jupiter family comet 67P/Churyumov-Gerasimenko (Schuhmann et al. 2019). Although they are more difficult to detect spectrally, $N_2$ and CO are also likely to be present on those bodies (e.g., Tegler et al. 2008).

The saturation vapor pressure of a substance is the pressure at which the gas phase is in thermodynamic equilibrium with a condensed (solid or liquid) phase; condensation and sublimation (or evaporation) rates are equal under this condition. Knowledge of vapor pressures as a function of temperature for Solar System ices is crucial for modeling processes such as seasonal migration of planetary volatiles, loss of volatiles from planetesimals, cometary gas production, compositional evolution of gas and solids in protoplanetary nebulae, and thermally driven evolution of volatile-rich planetary landscapes.

Fray and Schmitt (2009) compiled a valuable compendium of knowledge of vapor pressures for 27 species, including $N_2$, CO, and $CH_4$. They reviewed the ensemble of available laboratory data and provided convenient polynomial expressions for vapor pressures as a function of temperature. However, the majority of available laboratory measurements they used were for higher temperatures than are relevant to outer Solar System environments. For instance, available data for CO were at 54.78 K and above, and for $CH_4$ at 48.15 K and above, compared with equilibrium temperatures for a low albedo sphere of 30, 25, and 22 K at Kuiper belt distances of 30, 45, and 60 AU from the Sun, respectively. The scarcity of lower temperature measurements partly owes to the fact that laboratory measurement of pressures is much more straightforward in a dense, collisional gas, since the collisions equalize pressure throughout the apparatus and allow the fluid to be treated as a continuum substance. However, when the mean free path of gas molecules is comparable to (Knudsen flow) or much larger than (free molecular flow) the spatial scale of the laboratory apparatus, pressure becomes a function of geometry and location within the apparatus (e.g., Loeb 1934). Furthermore, the gas can be in local disequilibrium with itself since it is collisions that would drive equilibration of thermodynamic properties such as pressure and temperature. The pressure measured in these regimes by a pressure gauge is thus not necessarily the same as the pressure of interest at the equilibrium interface between gaseous and condensed phases. The Clausius-Clapeyron relation can be used to extrapolate from higher temperature data, but generally requires an assumption about the temperature dependence in the latent heat of sublimation to simplify the problem. Direct measurement of vapor pressures at lower temperatures and pressures would thus be valuable for refining models of sublimation and condensation processes in the outer Solar System. Therefore, in this study we used a quartz crystal microbalance to measure sublimation mass flux of $N_2$, CO, and $CH_4$ in the free molecular flow regime, accounting for the simultaneous sublimation from and condensation onto the quartz crystal, which allowed us to derive vapor pressures and latent heats of sublimation for these materials.

# Methods

## *Experimental Setup*

The natural vibration frequency of a piezoelectric quartz crystal declines as mass adheres to it.  This property is exploited in a quartz crystal microbalance (QCM) and provides a sensitive way to measure the mass balance of ice condensation and sublimation (e.g., Sack & Baragiola 1993; Allodi et al. 2013; Luna et al. 2014, 2018; Hudson et al. 2022).  In the Astrophysical Materials Laboratory at Northern Arizona University, we have integrated a QCM operated by an Inficon® IC6 controller into a cryogenic system used for study of the infrared optical constants of volatile ices.  A custom-made copper QCM holder from MCVAC Manufacturing® is attached via a copper strap to the tip of an Advanced Research Systems® (ARS) DE-204PB two stage closed-cycle helium cold head capable of reaching temperatures below 10 K (Fig. 1).  The cold head descends from a rotatable stage so that the gold-plated QCM face can be oriented toward a ThermoElectron Nicolet® iS50 FTIR spectrometer for spectral measurements, or toward other instrument ports not used in this study.  The chamber is assembled from stainless steel components using ConFlat® flanges with copper gaskets.  We pump the vacuum chamber with a Varian Agilent TwisTorr® 305 FT turbomolecular pump backed by a Varian Agilent® DS 302 rotary vane roughing pump.

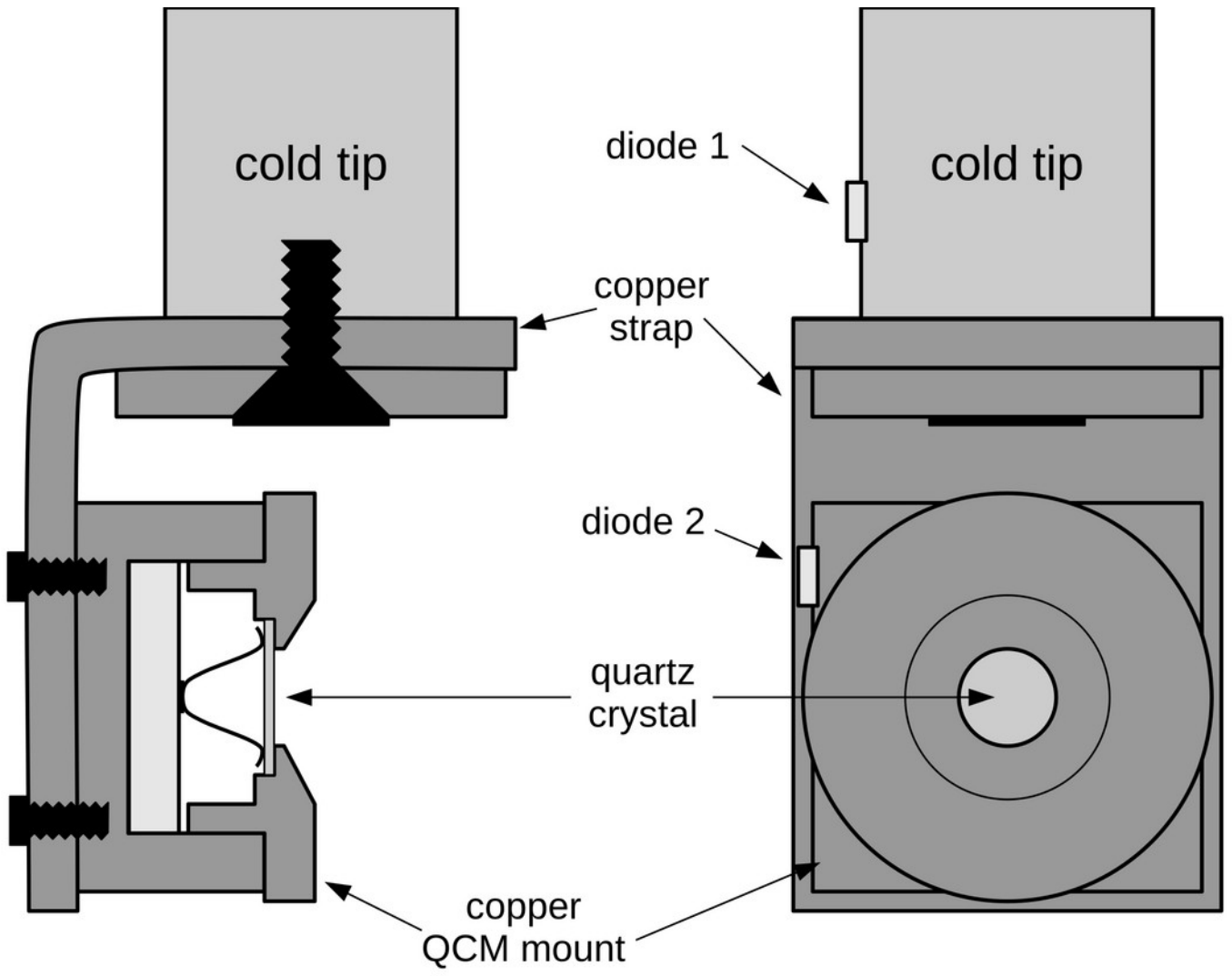

**Figure 1:** Schematic diagram of quartz crystal microbalance mount and attachment to cold tip.  Left, in cross section as viewed from the side.  Right, exterior view showing the face of the crystal where ice is deposited, with the placement of the two diode thermometers indicated.  For clarity, wiring to the diodes and to the crystal is omitted, as are the sheets of indium foil used to improve thermal conduction between adjacent copper parts.

Base pressure at room temperature is typically 1 to 2 × $10^{-8}$ torr, and well below $10^{-8}$ torr when the cryocooler is operating. We prepare materials to be studied as room temperature gases in an adjacent mixing manifold and then admit them to the chamber through an Agilent® model 951-5106 variable leak valve, with chamber pressures during deposition typically in the range from $10^{-7}$ to $10^{-5}$ torr. There is no line of sight between the leak valve and the QCM, which we oriented toward the chamber wall during our experiments. This geometry results in background deposition, where gas molecules arrive at the QCM from random directions after having first interacted with the room temperature chamber walls. We monitor the chamber pressure with an Inficon® BPG400 Bayard-Alpert Pirani combination gauge and also a Stanford Research Systems® RGA-200 quadrupole mass spectrometer (QMS). These two instruments are mounted on separate 1.5 inch diameter tubes extending off of the main chamber, with no line of sight to the QCM so they are primarily detecting gas molecules that last interacted with the chamber walls. They also have no line of sight to each other to minimize potential interference between the instruments. We measure the temperature with a pair of DT-670 diode thermometers, one affixed to the cold tip and the other to the copper QCM holder. We control the temperature with a Lake Shore® model 335 temperature controller that powers a 50 Ω heater wrapped around the cold tip.

## *Basic Equations*

We can compute the time rate of change in areal mass density $Q$ (mass $m$ per unit area $A$) on the QCM from the frequencies $f_1$ and $f_2$ measured at two times $t_1$ and $t_2$ according to the equation

$$\frac{1}{A}\frac{dm}{dt} = \frac{dQ}{dt} = \frac{c(f_1 - f_2)}{f_1 f_2 (t_2 - t_1)} \qquad \text{(eq. 1)}$$

where the constant $c$ is 4.417 × $10^5$ Hz g $cm^{-2}$ (Lu & Lewis 1972). Since the frequency of the QCM also depends on its temperature, this equation is only valid while the temperature is being held constant. During sublimation, the mass flux according to the Hertz-Knudsen-Langmuir equation (Langmuir 1913) is

$$\frac{dQ}{dt} = -p_{\mathrm{vap}}\sqrt{\frac{M}{2\pi R T_{\mathrm{QCM}}}} \qquad \text{(eq. 2)}$$

where $T_{\mathrm{QCM}}$ is the temperature of the QCM and its ice film, $p_{\mathrm{vap}}$ is the vapor pressure at that temperature, $M$ is the molecular mass, and $R$ is the gas constant. Inverting this equation to obtain $p_{\mathrm{vap}}$ is trivial, and is all that would be necessary for sublimation into a perfect vacuum.

However, no vacuum is perfect, especially not in a laboratory. While molecules are sublimating from the surface of the ice film, gas molecules from elsewhere in the apparatus are also arriving at and some of them are sticking to the ice, according to the analogous equation

$$\frac{dQ}{dt} = S_c\, p_{\mathrm{QCM}} \sqrt{\frac{M}{2\pi R T_{\mathrm{room}}}} \qquad \text{(eq. 3)}$$

where $p_{\mathrm{QCM}}$ is an effective gas pressure that accounts for sticking on the QCM (this is distinct from $p_{\mathrm{vap}}$) and $T_{\mathrm{room}}$ is the temperature of the chamber walls (room temperature) that sets the temperature of these impinging molecules. $S_c$ is a sticking coefficient that can generally be assumed to be unity for molecules arriving at a solid composed of like molecules at low temperature (Langmuir 1913). The nature of the substrate influences sticking, with softer crystal lattices favoring it (e.g., Leitch-Devlin & Williams 1985) so our assumption of unit sticking coefficients seems reasonable for weakly bonded solids such as $N_2$, CO, and $CH_4$. Bisshop et al. (2006) reported high sticking coefficients for CO and $N_2$ at 14 K and Brann et al. (2021) reported similar results for $CH_4$ at 20 K. In the free molecular flow regime where collisions between gas molecules are rare, gas fluxes onto and off of the ice are independent of one another and the mass flux measured by the QCM is simply the net of the sublimation and condensation terms (eqs. 2 and 3).

In the free molecular flow regime, $p_{\mathrm{QCM}}$ cannot simply be measured with a pressure gauge mounted somewhere in the chamber. However, at temperatures low enough that sublimation is negligible, we can record $p_{\mathrm{gauge}}$ at the same time as we record the condensation mass flux $dQ/dt$ onto the QCM, which provides an indirect measurement of $p_{\mathrm{QCM}}$, since we know the other parameters in eq. 3. From these two pressure measurements we can compute a correction factor $\Phi$ defined as the ratio of the two, $\Phi \equiv p_{\mathrm{QCM}}/p_{\mathrm{gauge}}$. This $\Phi$ factor depends on and can be used to correct for pressure gauge effects such as nonlinearity and species-dependent ionization efficiency as well as geometric effects associated with the configuration of the apparatus in the free molecular flow regime.

## *Initial Tests*

Prior to measuring vapor pressures, we performed tests to see if $\Phi$ has any dependence on temperature or pressure. To test for temperature dependence, we did a series of depositions at 10, 20, 30, 40, 50, and 60 K using ethane at a constant chamber pressure of $2 \times 10^{-6}$ torr. We selected ethane for this test since it is stable against sublimation over a broader temperature range than the hypervolatiles $N_2$, CO, and $CH_4$. We used ethane from Matheson® with purity 99.99 %. The areal deposition rate was approximately the same at each of these temperatures ($1.26 \times 10^{-8}$ g cm$^{-2}$ s$^{-1}$), corresponding to $\Phi$ values of 0.35 with a variance of ±1 %.

To test for pressure dependence, we performed depositions of $N_2$, CO, and $CH_4$ at a series of pressures between $10^{-8}$ and $10^{-4}$ torr while holding the QCM temperature constant at 10 K. For this and subsequent experiments described in this paper, we used research grade $N_2$ from Praxair® with purity 99.999 %, CO from Air Liquide® with purity 99.99 %, and $CH_4$ from Airgas® with purity 99.999 %. The results are shown in Fig. 2, revealing distinct $\Phi$ values for each species due to their different ionization efficiencies. The pressure dependence of $\Phi$ looks qualitatively similar for all three species, with a steep decline toward low pressures that we attribute to residual non-condensable gas in the system. According to the mass spectrometer, hydrogen is the most abundant of these by far, though helium or neon would behave the same. Hydrogen contributes to the pressure recorded in $p_{\mathrm{gauge}}$ but does not condense on the QCM. At higher pressures, we see a weak linear dependence on the log of pressure for which we do not have an explanation. These curves are reproducible from one day to the next, but they do depend on the orientation of the QCM, so it is important to measure $\Phi$ in the same geometry as will be used for the sublimation experiment.

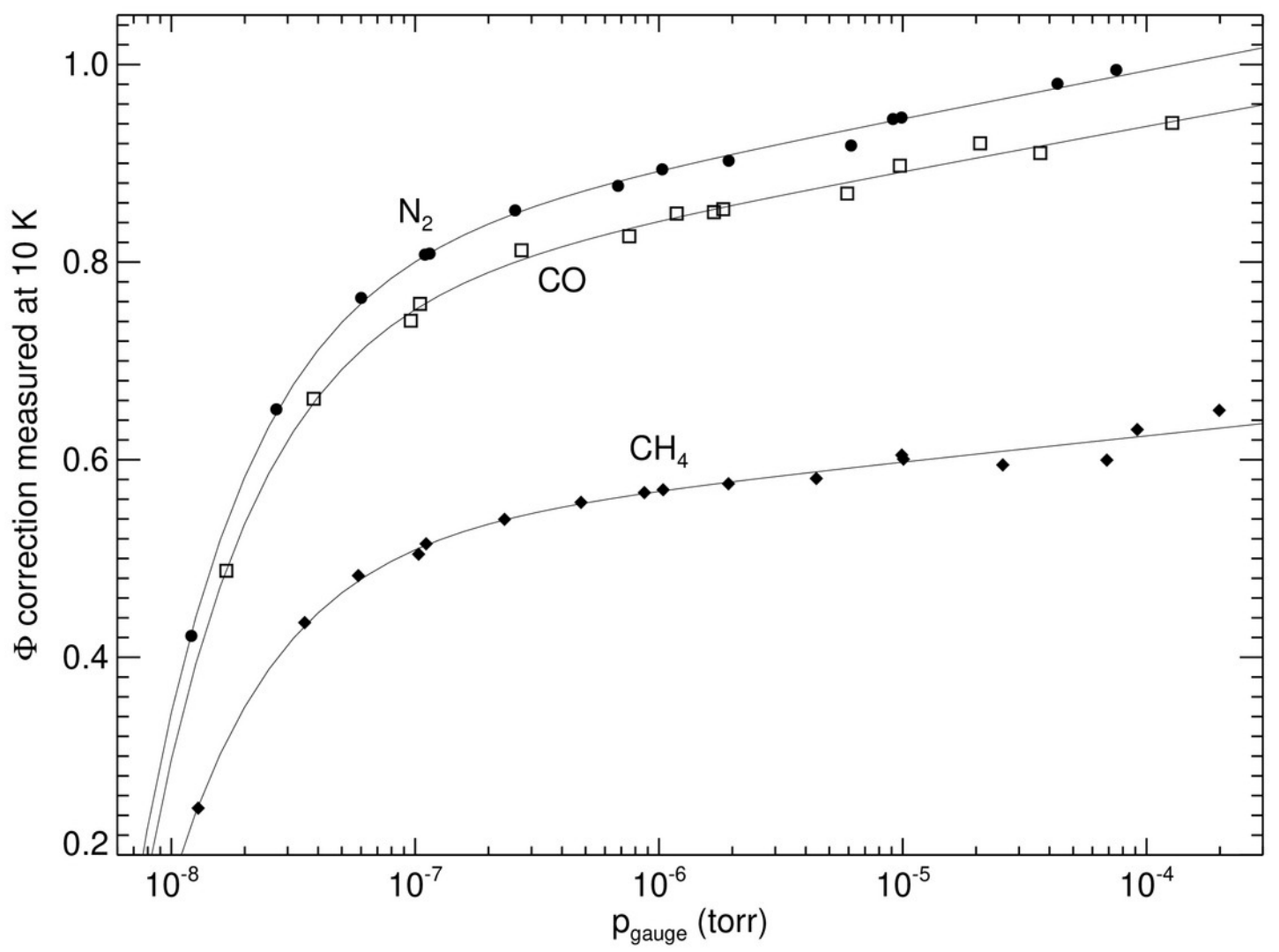

**Figure 2:** The correction factor $\Phi \equiv p_{\mathrm{QCM}}/p_{\mathrm{gauge}}$ accounts for the difference between pressure measured by the gauge ($p_{\mathrm{gauge}}$) and effective pressure at the QCM as measured by condensation onto the QCM. The points are measurements from deposition at various gauge pressures, with $T_{\mathrm{QCM}}$ held constant at 10 K. The smooth curves are a model consisting of a linear function in log($p$) modified by the effect of non-condensable contaminants at low pressures, as described by eq. 4. The fitted parameters $c_0$, $c_1$, and $p_0$ for the $N_2$ experiment were 1.1875, 0.0484, and $5.7 \times 10^{-9}$ torr, respectively. For CO they were 1.1200, 0.0456, and $6.1 \times 10^{-9}$ torr, and for $CH_4$ they were 0.7290, 0.0262, and $6.7 \times 10^{-9}$ torr.

## *Experimental Procedure*

The first step in a sublimation experiment is to deposit ice at a series of pressures between $10^{-8}$ and $10^{-5}$ torr to obtain a set of $\Phi$ values as in Fig. 2 and then fit a three parameter model consisting of a linear function in log($p$) scaled by the effect of residual non-condensable contaminants such as hydrogen

$$\Phi_{\mathrm{model}} = \left(c_0 + c_1 \log(p)\right)\left(\frac{p_{\mathrm{gauge}} - p_0}{p_{\mathrm{gauge}}}\right) \tag{eq. 4}$$

where the fitted parameters $c_0$, $c_1$, and $p_0$ are the constant and linear coefficients and the background partial pressure of non-condensable contaminants, respectively. This function is used to convert $p_{\mathrm{gauge}}$ to $p_{\mathrm{QCM}}$ during the sublimation part of the experiment. A more sophisticated scheme not developed here could be to monitor distinct contaminant species using the mass spectrometer and account separately for those that are condensable (such as $H_2O$ and $CO_2$) and those that are non-condensable (such as $H_2$ and, at higher temperatures, $N_2$). Rather than being a simple function of $p_{\mathrm{gauge}}$ and composition, $\Phi$ would then become a more general correction factor accounting for condensation onto the QCM as a function of measured partial pressures of all relevant species.

After depositing an ice film, we turn off the closed-cycle helium refrigerator and allow the cold head to warm passively for several minutes until the chamber pressure rises above ~$10^{-5}$ torr. We do this to sublimate away excess ice condensed elsewhere on the cold head, which enables us to more quickly achieve lower background pressures during the rest of the experiment. After turning the refrigerator back on, we control to a temperature where net sublimation occurs. When the QCM temperature has stabilized (about a minute after the diode thermometers show stable temperatures) and $p_{\mathrm{gauge}}$ has also stabilized (which can take a few minutes, or much longer if we had not driven off excess ice from the cold head by turning off the refrigerator earlier), we record the QCM frequency simultaneously with $p_{\mathrm{gauge}}$ for a few minutes. We then change to another temperature, and repeat until the ice sample has completely sublimated away. Preliminary tests showed that ice deposits much thicker than ~10 µm produced less consistent results, so we limited our ice films to a maximum of 5 µm thickness and performed multiple deposition and sublimation cycles to collect data at higher temperatures where the ice is rapidly lost to sublimation. To minimize the effect of contaminants, we limit our measurement of sublimation to chamber pressures above ~$10^{-7}$ torr where the species being studied dominates the composition of gas in the chamber.

During an interval when the temperature and pressure are deemed to be stable, we fit a line to the recorded QCM frequency versus time to obtain initial and final frequencies $f_1$ and $f_2$. These values are plugged into eq. 1 along with the initial and final times $t_1$ and $t_2$ to get the change in areal mass density as a function of time $dQ/dt$, which should equal the sum of the sublimation and condensation terms expressed in eqs. 2 and 3. Combining those equations, we have

$$\frac{dQ}{dt} = \left(\Phi\, p_{\text{gauge}} T_{\text{room}}^{-1/2} - p_{\text{vap}} T_{\text{QCM}}^{-1/2}\right)\sqrt{\frac{M}{2\pi R}} \ . \qquad \text{(eq. 5)}$$

Eq. 5 can be rearranged to solve for $p_{\text{vap}}$ at temperature $T_{\text{QCM}}$ as a function of measured and known quantities:

$$p_{\text{vap}} = \Phi\, p_{\text{gauge}}\sqrt{\frac{T_{\text{QCM}}}{T_{\text{room}}}} - \frac{dQ}{dt}\sqrt{\frac{2\pi R T_{\text{QCM}}}{M}} \ . \qquad \text{(eq. 6)}$$

We repeated our vapor pressure measurements multiple times at a few specific temperatures (36, 38, and 40 K for $CH_4$; 30 K for CO; and 27 K for $N_2$) using various ice thicknesses up to 5 µm and thermal histories (warming and cooling sequences) to estimate our measurement uncertainty. The standard deviations of $p_{\text{vap}}$ from these tests indicate vapor pressure uncertainties of ±9 %. An additional uncertainty comes from the calibration of the temperature measurement. We used two phase changes that occur at known temperatures to refine our temperature calibration: the solid-solid phase change between $CH_4$ I and $CH_4$ II at 20.4 K and the solid-liquid phase change in $C_3H_8$ at 85.5 K. These tests are described in greater detail in Appendix A. After correcting for a linear interpolation between the observed offsets between measured and known temperatures of these two transitions, we estimate a residual temperature uncertainty of about ±0.2 K.

## Results and Discussion

Fig. 3 shows vapor pressures $p_{\text{vap}}$ as a function of temperature for α $N_2$, α CO, and $CH_4$ I determined via the methodology described in the previous section. We also plot the widely-used polynomial expressions from Fray & Schmitt (2009) as well as earlier curves from Brown & Ziegler (1980). For each of the three species, our points differ somewhat from the literature curves. The greatest discrepancy is seen for CO, where our vapor pressures are nearly an order of magnitude below the Fray & Schmitt recommended polynomial approximation. The Brown & Ziegler curve for CO is somewhat closer to our results, though for $N_2$, the Fray & Schmitt curve is closer. The Brown & Ziegler and Fray & Schmitt curves for $CH_4$ are the same.

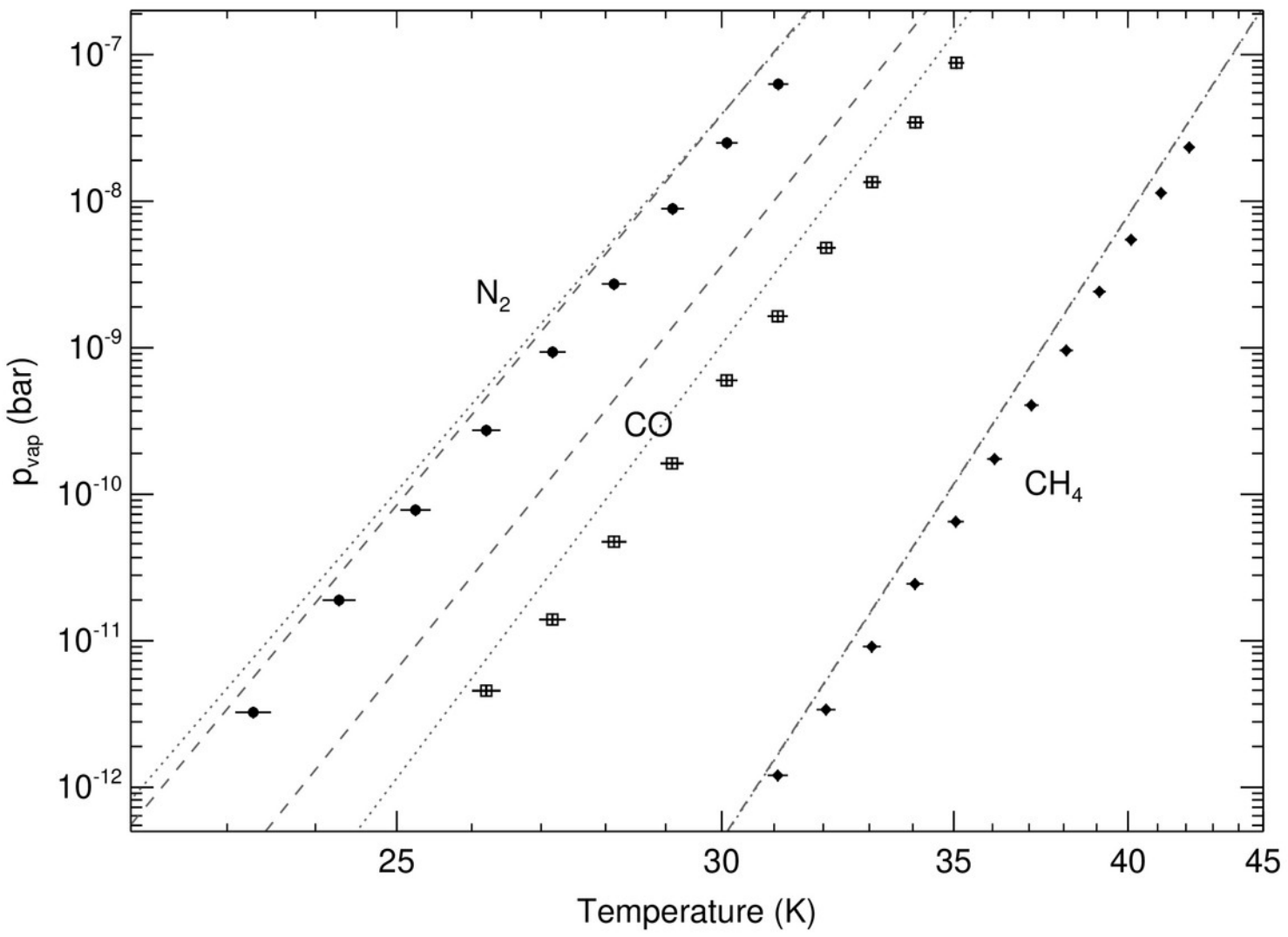

**Figure 3:** Our measured vapor pressures (points) compared with polynomial expressions (dashed curves from Fray & Schmitt 2009, dotted curves from Brown & Ziegler 1980). The measured temperatures have been corrected as described in Appendix A, and are plotted on a scale that is linear in 1/*T*. Vertical error bars represent ±9 % random errors in measured $p_{vap}$ while horizontal error bars represent ±0.2 K systematic uncertainty in measured temperature.

## *Latent Heat of Sublimation*

The Clausius-Clapeyron relation expresses the change in pressure with temperature as a function of pressure *p*, temperature *T*, and latent heat of sublimation *L*, assuming the volume of the solid is negligible compared to the volume of the gas and the gas behaves as an ideal gas:

$$\frac{dp}{dT} = \frac{p\,L}{T^2 R} \,. \qquad \text{(eq. 7)}$$

This equation can be rearranged to separate pressure and temperature terms and both sides integrated upon an assumption of constant *L* to obtain

$$\ln(p) = \ln(p_0) + \frac{L}{R}\left(\frac{1}{T_0} - \frac{1}{T}\right) , \qquad \text{(eq. 8)}$$

where $T_0$ and $p_0$ are constants corresponding to a reference temperature and pressure pair that lie along the phase change curve. Rather than select a single laboratory-measured pair (e.g., Steckloff et al. 2015) we arbitrarily set $T_0$ to a temperature in the middle of the range we measured for each of our species, and found the best-fitting values of $p_0$ and *L* for our full set of

**Table 1**

Constant Latent Heat

| Species | $T_0$ (K) | $p_0$ ($10^{-9}$ bar) | $L$ (kJ mol$^{-1}$) | $L_{NIST}$ (kJ mol$^{-1}$) |
|---|---|---|---|---|
| α $N_2$ | 27 | 0.76 ± 0.02 | 7.56 ± 0.07 | 7.34 |
| α CO | 31 | 1.64 ± 0.05 | 8.63 ± 0.08 | 7.60 |
| $CH_4$ I | 37 | 0.43 ± 0.01 | 9.81 ± 0.08 | 9.70 |

vapor pressure measurements for that species. The values thus obtained are shown in Table 1 along with uncertainties arising from the estimated ±9 % uncertainties in $p_{vap}$. These $L$ values can be compared to the National Institute of Standards and Technology (NIST) values in the last column of the table (Stephenson & Malanowski 1987; Shakeel et al. 2010). As with $p_{vap}$, the largest discrepancy is for CO.

Since the latent heat of sublimation could be a function of temperature, a better approach is to make use of the exact Clapeyron equation which expresses $dp/dT$ as $L/(T\Delta v)$ where $\Delta v$ is the change in volume. Note that $L$ and $\Delta v$ can be temperature-dependent properties. Lobo & Ferreira (2001) use this equation to expand $\ln(p)$ into a series of terms:

$$\ln(p) = A - \frac{B}{T} + C\ln(T) + \sum_{i=2}^{4} D_i T^{i-1} + \frac{E(T)}{T} p \ . \qquad \text{(eq. 9)}$$

An advantage of this approach is that the parameters $A$, $B$, $C$, $D_2$, $D_3$, $D_4$, and $E(T)$ each correspond to specific properties of the material. The expressions for the parameters in terms of material properties are given in detail in their paper (Lobo & Ferreira 2001). While our data do not support solving for so many parameters, $C$, $D_2$, $D_3$, and $D_4$, can be drawn from literature sources and $E(T)$ can be neglected for low pressures, so we can use our data to fit for just $A$ and $B$. We also include additional $p_{vap}$ measurements from the literature in addition to our measurements to extend the temperature range. These additional data are from Borovik et al. (1960) for the lower temperature α phase of $N_2$, from Morrison et al. (1968) and Shinoda (1969) for α CO, and from Tickner & Lossing (1951) and Armstrong et al. (1955) for the higher temperature phase $CH_4$ I. The parameters $C$, $D_2$, $D_3$, and $D_4$ are derived from the isobaric heat capacities of solid phase and ideal gas. Heat capacities of solid phase are taken from Clayton & Giauque (1932), Leah (1956), Colwell et al. (1963), and Gavrilko et al. (1999), while that of ideal gas is $7R/2$ for $N_2$ and CO, and $4R$ for $CH_4$. The values of these parameters are listed in Table 2 and the data and fits are shown in Figs. 4-6.

**Table 2**

Parameters for Lobo & Ferreira expression

| Parameter [units] | α $N_2$ | β $N_2$ | α CO | β CO | $CH_4$ I |
|---|---|---|---|---|---|
| $A$ | **−3.92 ± 0.21** | 9.4928 | **−5.963 ± 0.036** | −49.4956 | **−2.510 ± 0.027** |
| $B$ [K] | **805.3 ± 5.3** | 857.5053 | **982.5 ± 2.1** | 606.5968 | **1139.3 ± 2.0** |
| $C$ | 4.54794 | 1.00304 | 6.32108 | 17.15769 | 3.85295 |
| $D_2$ [$K^{-1}$] | −0.08002 | −0.04581 | −0.16588 | −0.27704 | −0.06385 |
| $D_3$ [$K^{-2}$] | $-5.347 \times 10^{-5}$ | $2.630 \times 10^{-4}$ | $1.068 \times 10^{-3}$ | $6.394 \times 10^{-4}$ | $2.281 \times 10^{-4}$ |
| $D_4$ [$K^{-3}$] | 0 | $-1.210 \times 10^{-6}$ | $-5.00 \times 10^{-6}$ | 0 | $-5.42 \times 10^{-7}$ |

Table note: bold figures are our fitted values for $A$ and $B$ along with their uncertainties $\sigma_A$ and $\sigma_B$. Other figures are based on literature values as described in the text.

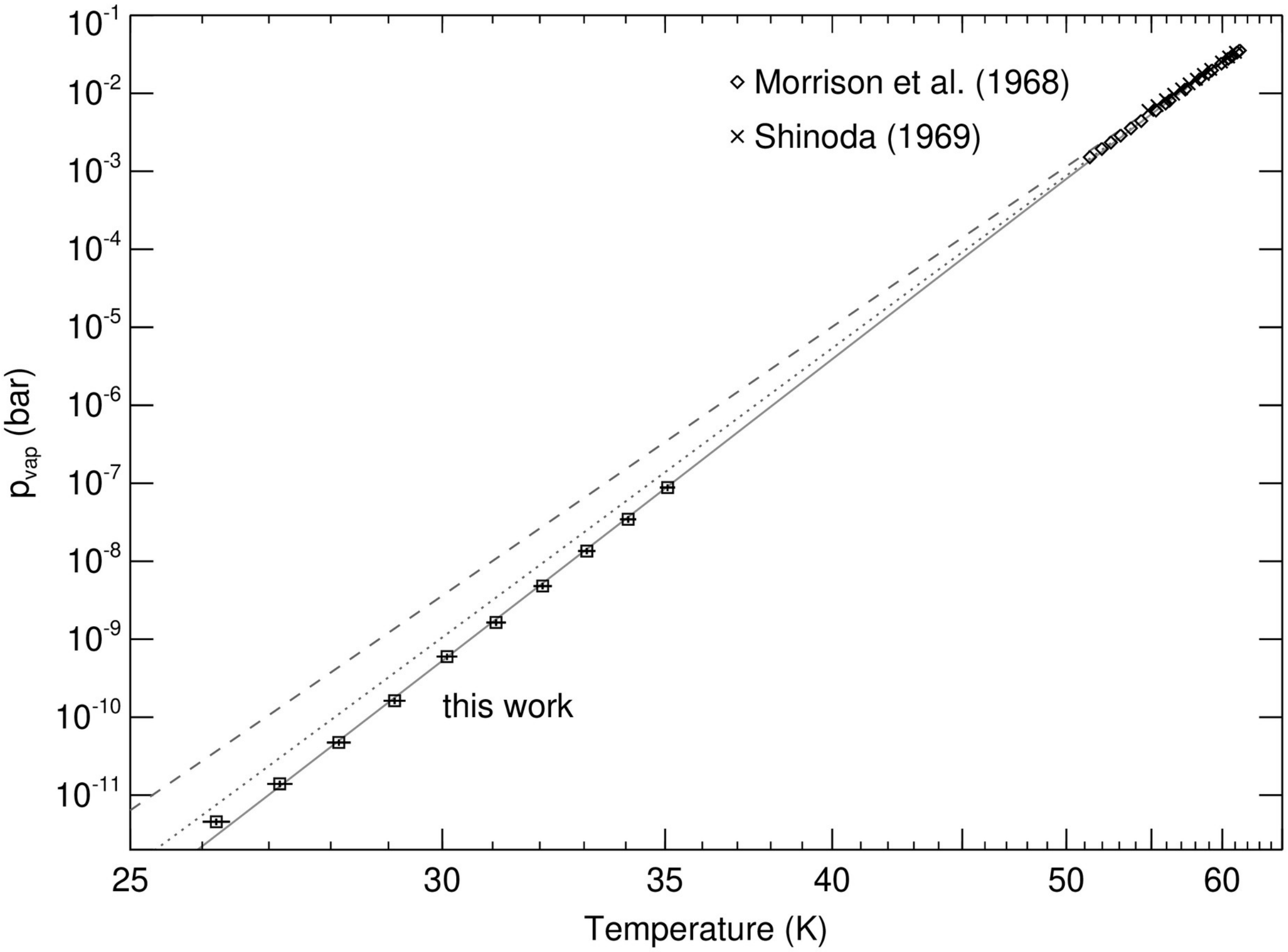

**Figure 4:** Our α CO data (open squares) compared with data from Morrison et al. (1968, open diamonds), Shinoda (1969, × symbols) and polynomials from Fray & Schmitt (2009, dashed line) and Brown & Ziegler (1980, dotted line). The solid line is our version of the Lobo & Ferreira (2001) expression with $A$ and $B$ fitted to the combined data and $C$, $D_2$, $D_3$, and $D_4$ taken from the literature, with values shown in Table 2.

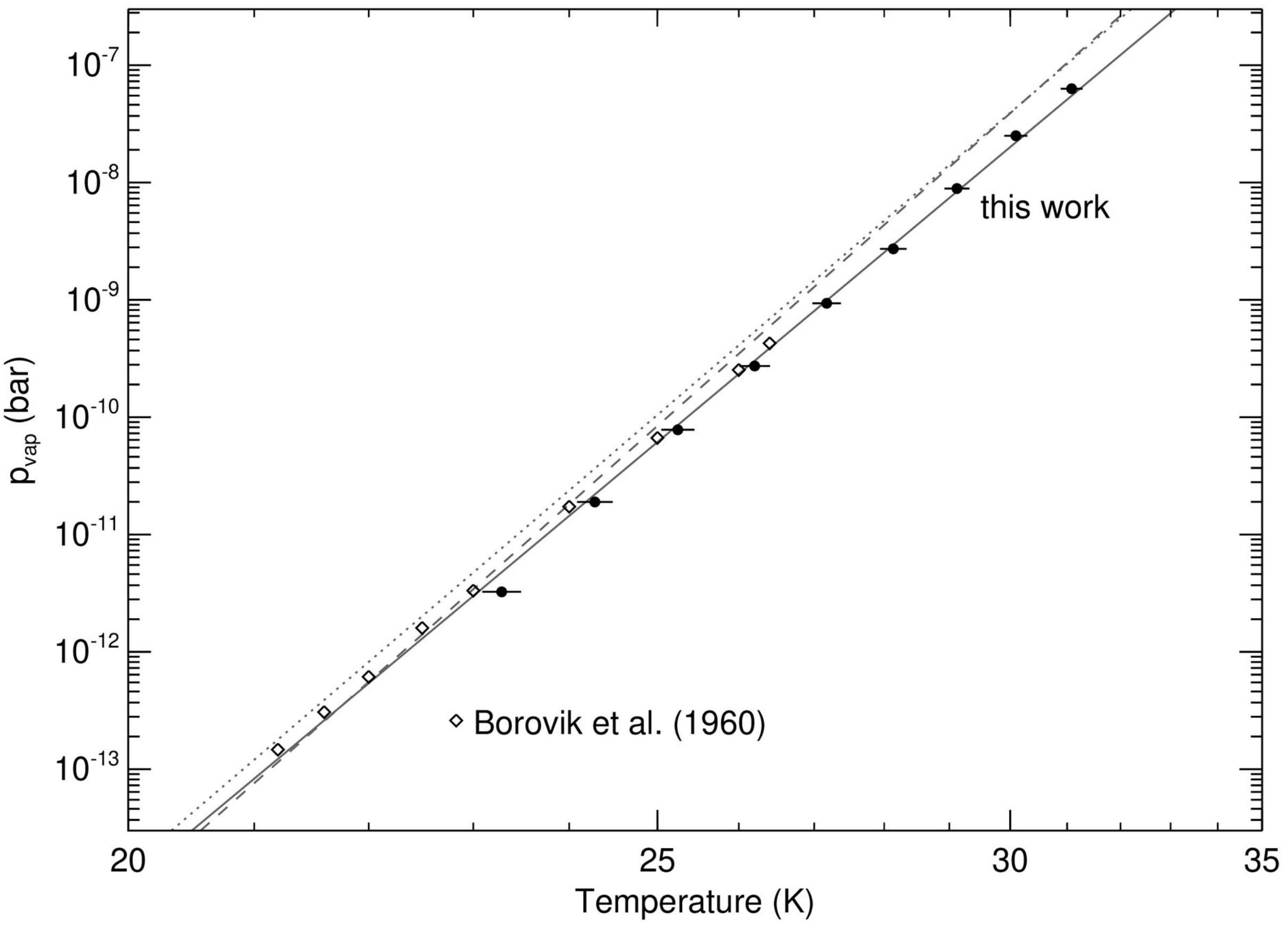

**Figure 5:** Our α $N_2$ data (black circles) compared with data from Borovik et al. (1960, open diamonds) and polynomials from Fray & Schmitt (2009, dashed line) and Brown & Ziegler (1980, dotted line). The solid line is our version of the Lobo & Ferreira (2001) expression with $A$ and $B$ fitted to the combined data and $C$, $D_2$, $D_3$, and $D_4$ taken from the literature, with values shown in Table 2.

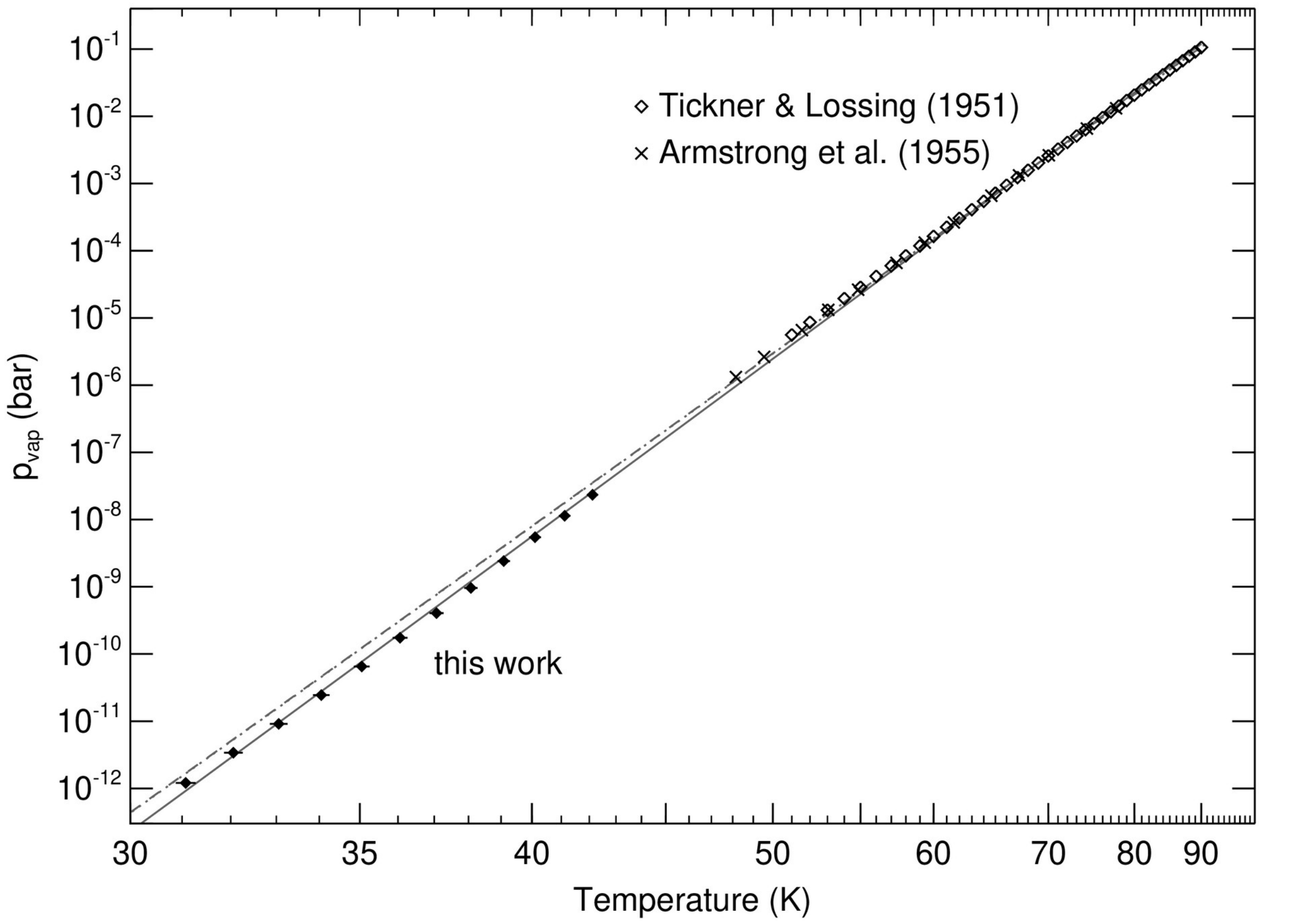

**Figure 6:** Our $CH_4$ I data (black diamonds) compared with data from Tickner & Lossing (1951, open diamonds), Armstrong et al. (1955, × symbols) and polynomials from Fray & Schmitt (2009, dashed line) and Brown & Ziegler (1980, dotted line). The solid line is our version of the Lobo & Ferreira (2001) expression with *A* and *B* fitted to the combined data and *C*, $D_2$, $D_3$, and $D_4$ taken from the literature, with values shown in Table 2.

We can retrieve the latent heat of sublimation as a function of temperature from the parameters *B*, *C*, $D_2$, $D_3$, $D_4$, according to the equation (Lobo & Ferreira 2001):

$$L(T) = R\left[B + C\,T + \sum_{i=2}^{4}(i-1)D_i T^i\right] \pm \sigma_B R \ . \qquad \text{(eq. 10)}$$

These functions are plotted in Fig. 7, along with values for β $N_2$ and β CO based on literature values.

## Data Availability

All laboratory data that we used to derive the results presented in this paper are publicly archived along with the values plotted in the figures at https://openknowledge.nau.edu/id/eprint/6246. The data consist of time series ASCII files listing calibrated temperature in K, gauge pressure in torr, and QCM frequency in Hz.  Values shown in the figures include our measurements of $\Phi$ versus pressure, vapor pressure versus temperature, and heat of sublimation versus pressure, including both measured discrete values plotted as points and the polynomials plotted as smooth curves.

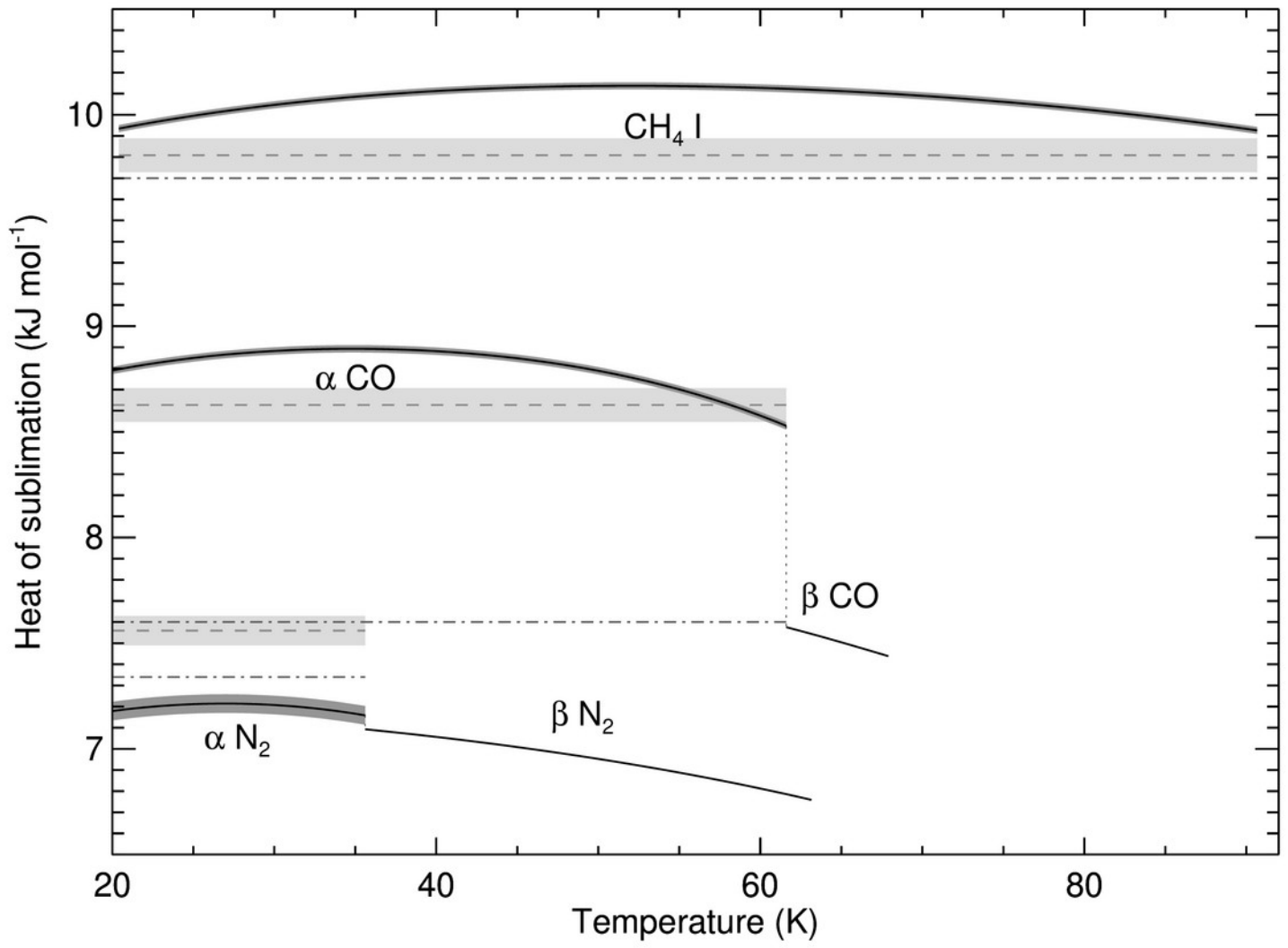

**Figure 7:** Solid black curves show temperature-dependent latent heats of sublimation from eq. 10, using the values of $B$, $C$, $D_2$, $D_3$, and $D_4$ in Table 2.  Adjacent gray areas indicate the uncertainty from $\sigma_B R$.  Dashed gray lines with adjacent gray areas representing uncertainties are our constant latent heat values from eq. 8 and Table 1.  Dot-dashed gray lines are the NIST values from Stephenson & Malanowski (1987) and Shakeel et al. (2018).

## *Astrophysical Implications*

The lower vapor pressures for these three volatile ices, and especially of CO, have important implications across a variety of astrophysical and planetary environments.  For instance, the CO snow line in a protoplanetary nebula will occur at a higher temperature and thus closer to the star than previously thought.  This means more planetesimals will be able to incorporate CO ice among their solid constituents, and then they will retain it for longer than they would have if the higher literature vapor pressures held (e.g., Umurhan & Birch 2023).  An order of magnitude lower vapor pressure corresponds to an order of magnitude lower sublimation rate at a given temperature and an order of magnitude longer retention time, all other things being equal.  On icy dwarf planets hosting gravitationally bound $N_2$ and CO, such as Triton, Pluto, and possibly Eris and Makemake, it will be easier for seasonal volatile transport processes to segregate $N_2$ from CO ices via solid state distillation, owing to their greater difference in volatility.  CO-driven activity on interstellar and Oort cloud comets and Centaurs will turn on closer to the Sun than otherwise would have been expected.  The higher latent heat values we derive for CO mean that approximately 10 % more energy is required to sublimate a given quantity of CO ice than what would be assumed based on the NIST value, and that CO is better able to transport heat through sublimation followed by condensation elsewhere.

## Acknowledgments

Portions of this work and laboratory facility were supported by NASA Solar System Workings Program grant 80NSSC19K0556; by the State of Arizona Technology and Research Initiative Fund (TRIF), administered by the Arizona Board of Regents; by NSF Research Experience for Undergraduates grant 1950901 to Northern Arizona University; and by philanthropic donations from the John and Maureen Hendricks Foundation and from Lowell Observatory's Slipher Society.

We thank two anonymous reviewers for their many helpful suggestions to improve this manuscript and also the free and open source software communities for empowering us with key tools used to complete this project, notably Linux, the GNU tools, LibreOffice, Evolution, Python, the Astronomy Users Library, and FVWM.

## Appendix A

Temperature calibration is crucial for these measurements, since $p_{vap}$ is highly sensitive to temperature. Our temperatures are measured with a pair of DT-670 diode thermometers, one affixed to the copper quartz crystal microbalance holder and one to the cold tip as shown in Fig. 1. Since the copper strap provides excellent thermal conductivity between the two diodes, and we have no reason to favor the calibration of one diode over the other, all measured temperatures are the average of the two. Across the range of temperatures from 10 to 100 K, the two diodes consistently report temperatures within 0.3 K of one another. The fact that they do not report identical temperatures supports averaging the two together to dilute idiosyncrasies of either one, and also calls for additional checks to minimize potential systematic temperature calibration issues, such as differences between the temperature measured by the diodes and the temperature of the QCM.

We exploited two phase changes that occur at known temperatures to refine our temperature calibration. The first is a solid-solid phase change that occurs at 20.4 K in $CH_4$ ice between warmer $CH_4$ I and the more ordered, lower temperature $CH_4$ II phase (e.g., Clusius 1929; Bol'shutkin et al. 1972). To pinpoint the temperature of the $CH_4$ I-II transition in our system, we performed a series of stepwise cooling sequences across the phase transition, collecting FTIR spectra of thin films of $CH_4$ ice at each temperature step after holding at that temperature for at least two minutes for the temperature to stabilize. The $\upsilon_4$, $\upsilon_3$, and $\upsilon_3+\upsilon_4$ vibrational absorption features of $CH_4$ at 1300, 3010, and 4300 $cm^{-1}$, respectively, all exhibit changes in shape and location between the two phases, as shown in Fig. A1. At each temperature step, we converted the FTIR spectrum to absorbance and computed the centroid frequency and the mean deviation from the centroid of each of the $CH_4$ bands. We plotted the rate of change with temperature of these six parameters versus our measured temperatures, finding the changes in all six exhibit narrow peaks at a measured temperature of 20.2 K which we take to correspond to the phase change temperature of 20.4 K.

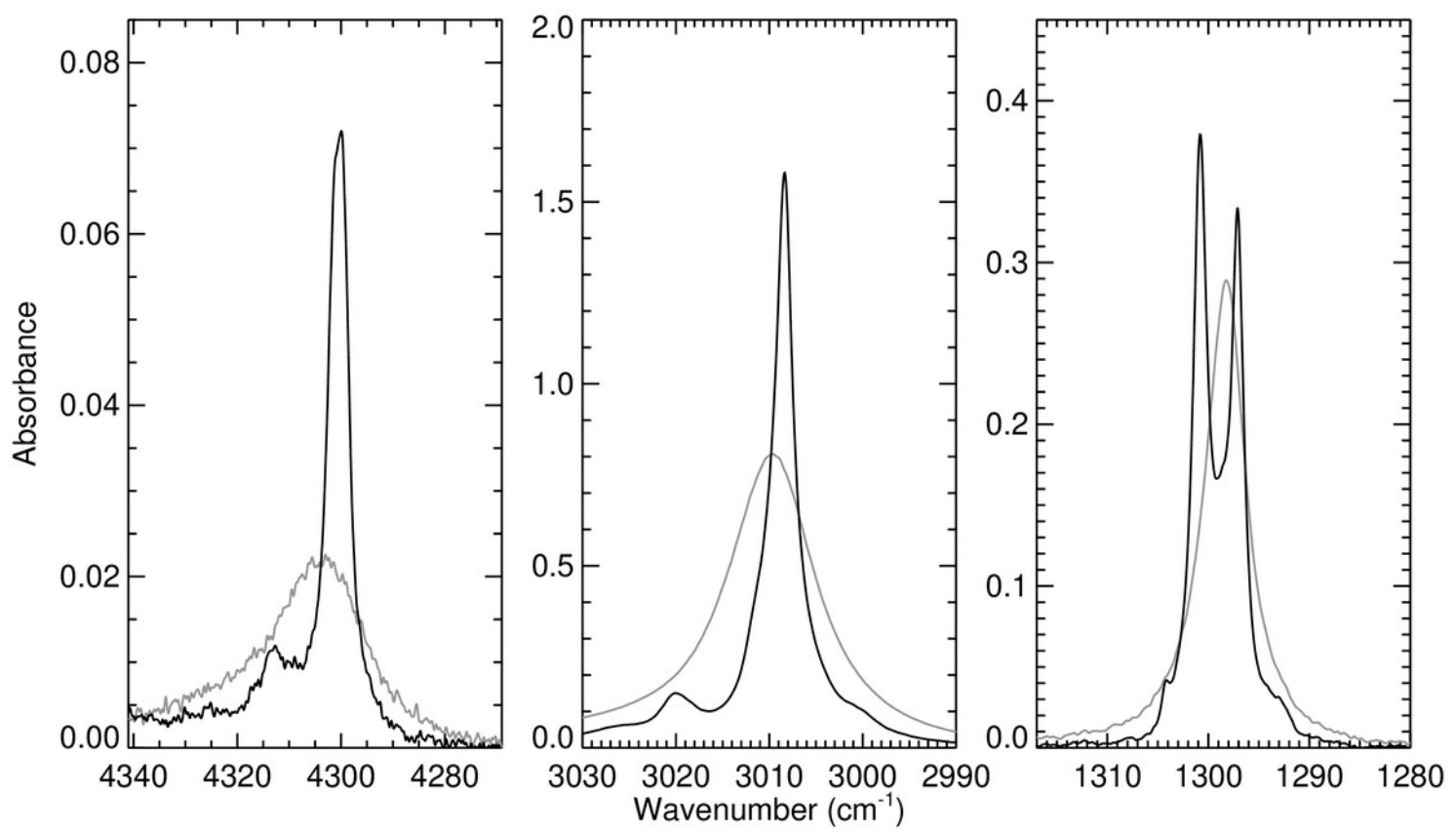

**Figure A1:** Comparison of absorption bands in $CH_4$ II at 15 K (black) and $CH_4$ I at 25 K (gray).

The second phase change we used for temperature calibration is the melting point of $C_3H_8$ at 85.45 to 85.50 K (e.g., Kemp & Egan 1938; Pavese & Besley 1981; Acree 1991; Perkins et al. 2009). This transition is readily detected using the QCM, since liquid $C_3H_8$ does not reduce the vibration frequency of the quartz crystal the way a coating of solid $C_3H_8$ does. Using 99.9974 % pure research grade $C_3H_8$ from Gas Innovations®, we performed a warming series through the phase change. We increased the temperature in 0.05 K steps, giving the QCM a few minutes at each step to ensure the temperature had stabilized. $C_3H_8$ melting was observed to occur between measured temperatures of 85.54 and 85.59 K, about 0.10 to 0.15 K off from the literature values, as shown in Fig. A2. Cooling again immediately after it melted revealed that at least some $C_3H_8$ remained on the QCM as a liquid, since an abrupt drop in frequency was observed when it re-froze.

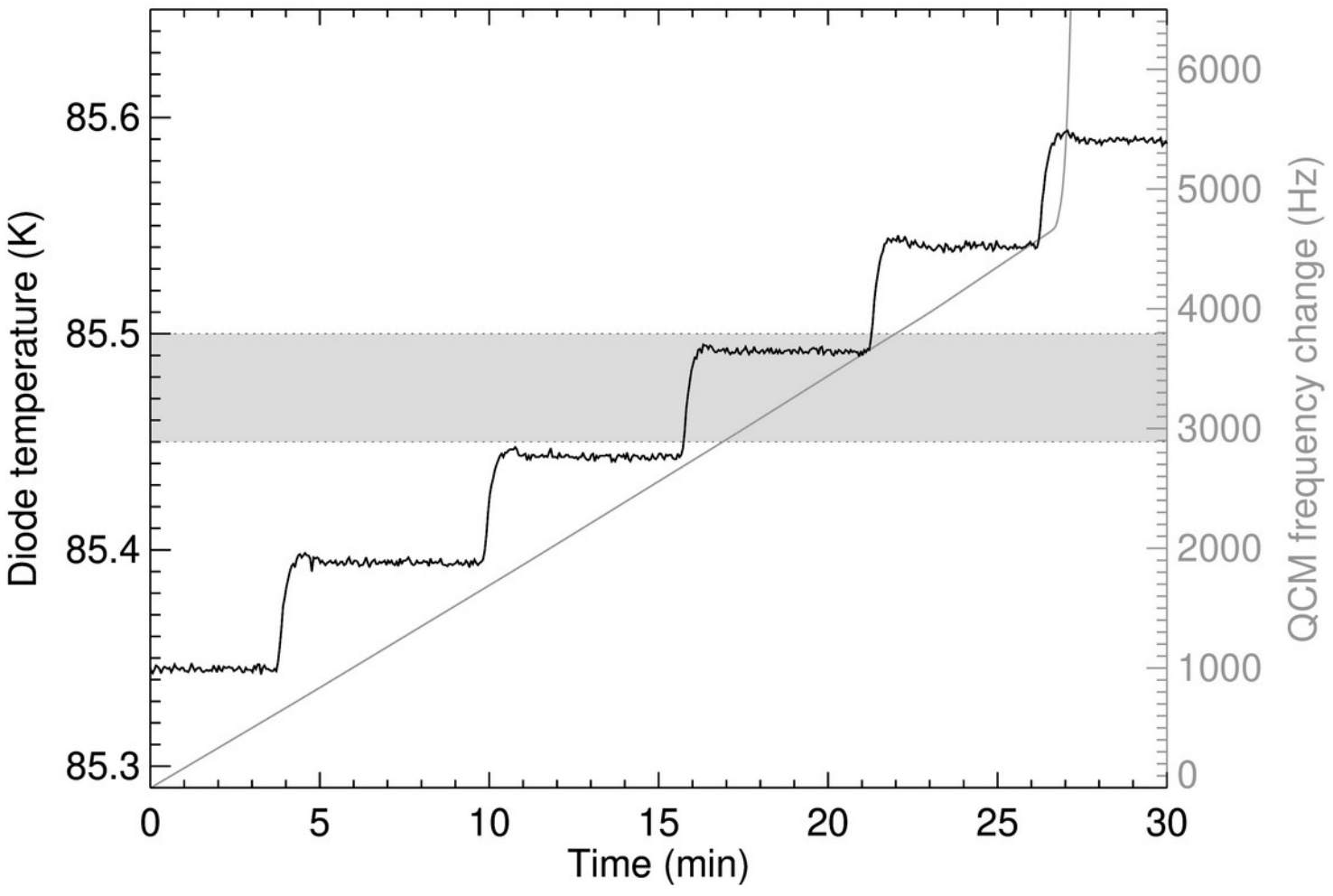

**Figure A2:** Temperature calibration warming sequence for $C_3H_8$. Black curve shows the temperature history (left axis), with melting expected in the gray horizontal band (Kemp & Egan 1938; Pavese & Besley 1981; Acree 1991; Perkins et al. 2009). The QCM frequency history (gray curve, right axis) shows that melting occurred soon after we increased the temperature from 85.54 K to 85.59 K, indicating that our diode temperatures read about 0.10 to 0.15 K too warm in this temperature range.

We computed a linear correction factor from our two temperature measurements by assuming the literature values were the true temperatures. We applied this correction to all of our measured temperatures, resulting in temperatures that are slightly shifted relative to the integer values we had controlled to. We estimate our temperature uncertainty to be about ±0.2 K, since we are unsure if the trend between the −0.2 K temperature offset observed at 20.4 K and the +0.1 K offset observed at 85.5 K is really linear.

We should note that many QCM systems are sold with stainless steel mounts. At low temperatures, stainless steel is a much worse conductor of heat than copper is. The small amount of power used to drive the vibration of the quartz crystal heats the crystal, and most of that heat is conducted away through the face plate of the QCM mount. The heating produces a discrepancy between the temperature of the quartz crystal itself and the temperature of the surface the QCM mount is affixed to. The discrepancy is greatest at low temperatures. During early tests with a stainless face plate, we observed the 20.4 K $CH_4$ I-II transition to occur at a diode temperature of about 12 K, indicating a temperature discrepancy of more than 8 K between the quartz crystal and the cold tip. Use of the much higher conductivity copper QCM mount and face plate reduced that discrepancy to only 0.2 K.